\documentclass[12pt]{article}

\usepackage{times}
\usepackage{amsfonts}

  \topmargin - 0.5 cm 
\begin{document}

\baselineskip= 0.80 true cm

\begin{center}
{\Large A UNIFIED APPROACH TO SIC-POVMs AND MUBs}

\bigskip

Olivier Albouy and Maurice R. Kibler

\bigskip

\textit{Universit\'{e} de Lyon, Institut de Physique Nucl\'{e}aire, \\[0pt]
Universit\'{e} Lyon 1 and CNRS/IN2P3, 43 bd du 11 novembre 1918, \\[0pt]
F--69622 Villeurbanne, France}

Electronic mail: o.albouy@ipnl.in2p3.fr, m.kibler@ipnl.in2p3.fr

\bigskip

\textbf{Abstract}
\end{center}

\begin{quotation}
A unified approach to (symmetric informationally complete) positive operator
valued measures and mutually unbiased bases is developed in this article.
The approach is based on the use of Racah unit tensors for the Wigner-Racah
algebra of $\mathrm{SU}(2)\supset \mathrm{U}(1)$. Emphasis is put on
similarities and differences between SIC-POVMs and MUBs.
\end{quotation}

\bigskip



\section{INTRODUCTION}

The importance of finite--dimensional spaces for quantum mechanics is well
recognized (see for instance \cite{01Peres}-\cite{add Muynck}). In 
particular, such spaces play a major role in quantum information theory, 
especially for quantum cryptography and quantum state tomography 
\cite{03Jauch}-\cite{21Albouy}. Along this vein, 
a symmetric informationally complete (SIC)
positive operator valued measure (POVM) is a set of operators acting on a
finite Hilbert space \cite{03Jauch}-\cite{11Weigert} (see also 
\cite{add Muynck} for an infinite Hilbert space) and mutually
unbiased bases (MUBs) are specific bases for such a space 
\cite{12Schwinger}-\cite{21Albouy}.

The introduction of POVMs goes back to the seventies 
\cite{03Jauch}-\cite{add Kraus}. The most general 
quantum measurement is represented by a POVM.
In the present work, we will be interested in SIC-POVMs, for which the
statistics of the measurement allows the reconstruction of the quantum
state. Moreover, those POVMs are endowed with an extra symmetry condition
(see definition in Sec.~2). The notion of MUBs (see definition in Sec.~3),
implicit or explicit in the seminal 
works of \cite{12Schwinger}-\cite{15Wootters}, has been the object of 
numerous mathematical and physical
investigations during the last two decades in connection with the so-called
complementary observables. Unfortunately, the question to know, for a given
Hilbert space of finite dimension $d$, whether there exist SIC-POVMs and how
many MUBs there exist has remained an open one.

The aim of this note is to develop a unified approach to SIC-POVMs and MUBs
based on a complex vector space of higher dimension, viz. $d^{2}$\ instead
of $d$. We then give a specific example of this approach grounded on the
Wigner-Racah algebra of the chain $\mathrm{SU}(2)\supset \mathrm{U}(1)$
recently used for a study of entanglement of rotationally invariant spin
systems \cite{22Breuer} and for an angular momentum study of MUBs 
\cite{20Kibler Planat,21Albouy}.

Most of the notations in this work are standard. Let us simply mention that 
$\mathbb{I}$ is the identity operator, the bar indicates complex
conjugation, $A^{\dagger}$ denotes the adjoint of the operator $A$, 
$\delta _{a,b}$ stands for the Kronecker symbol for $a$ and 
$b$, and $\Delta (a,b,c)$ is $1$ or $0$ according as $a$, $b$ and $c$ satisfy
or not the triangular inequality.

\section{SIC-POVMs}

Let $\mathbb{C}^{d}$\ be the standard Hilbert space of dimension $d$\
endowed with its usual inner product denoted by $\langle \ |\ \rangle $. As
is usual, we will identify a POVM with a nonorthogonal decomposition of the
identity. Thus, a discrete SIC-POVM is a set 
$\{P_{x} : x = 1, 2, \cdots, d^{2}\}$ of $d^{2}$ nonnegative operators $P_{x}$
acting on $\mathbb{C}^{d}$, such that:

\begin{itemize}
\item they satisfy the \textit{trace} or \textit{symmetry condition}
\begin{equation}
{\rm Tr}\left( P_{x}P_{y}\right) =\frac{1}{d+1},\quad x\not=y;
\label{Symmetry condition}
\end{equation}

moreover, we will assume the operators $P_{x}$ are 
normalized, thus completing this condition with
\begin{equation}
{\rm Tr}\left( P_{x}^{2}\right) =1;
\label{normalizationcondition}
\end{equation}

\item they form a \textit{decomposition of the identity}
\begin{equation}
\frac{1}{d}\sum_{x=1}^{d^{2}}P_{x}=\mathbb{I};
\end{equation}

\item they satisfy a \textit{completeness condition}: the knowledge of the 
probabilities 
$p_{x}$ defined by $p_{x}={\rm Tr}(P_{x}\rho )$\ is sufficient to
reconstruct the density matrix $\rho$.
\end{itemize}

\bigskip

Now, let us develop each of the operators $P_{x}$ on an orthonormal 
(with respect to the Hilbert--Schmidt product) basis 
$\{u_{i} : i = 1, 2, \cdots, d^{2}\}$ of the space of linear operators on 
$\mathbb{C}^{d}$
\begin{equation}
P_{x}=\sum_{i=1}^{d^{2}}v_{i}(x)u_{i},
\end{equation}
where the operators $u_{i}$ satisfy 
${\rm Tr}(u_{i}^{\dagger} u_{j}) = \delta_{i,j}$. The 
operators $P_{x}$ are thus considered as vectors
\begin{equation}
v(x)=(v_{1}(x), v_{2}(x), \cdots, v_{d^{2}}(x))
\end{equation}
in the Hilbert space $\mathbb{C}^{d^{2}}$\ of dimension $d^{2}$ and the
determination of the operators $P_{x}$ is equivalent to the determination of
the components $v_{i}(x)$ of $v(x)$. In this language, the trace property 
(\ref{Symmetry condition}) together with the normalization condition 
(\ref{normalizationcondition}) give
\begin{equation}
v(x)\cdot v(y)=\frac{1}{d+1}\left( d\delta _{x,y}+1\right) ,
\label{Result 1}
\end{equation}
where $v(x)\cdot v(y)=\sum_{i=1}^{d^{2}}\overline{v_{i}(x)}v_{i}(y)$\ is the usual Hermitian product in $\mathbb{C}^{d^{2}}
$.

In order to compare Eq.~(\ref{Result 1}) with what usually happens in the
search for SIC-POVMs, we suppose from now on that the operators $P_{x}$\ are
rank-one operators. Therefore, by putting 
\begin{equation}
P_{x}=|\Phi _{x}\rangle \langle \Phi _{x}|
\end{equation}
with $|\phi _{x}\rangle \in \mathbb{C}^{d}$, the trace property 
(\ref{Symmetry condition}, \ref{normalizationcondition})
reads
\begin{equation}
|\langle \Phi _{x}|\Phi _{y}\rangle |^{2}=\frac{1}{d+1}\left( d\delta
_{x,y}+1\right) .  \label{Trace property in phi}
\end{equation}
From this point of view, to find $d^{2}$ operators $P_{x}$ is equivalent to
finding $d^{2}$ vectors $|\phi _{x}\rangle $ in $\mathbb{C}^{d}$ satisfying
Eq.~(\ref{Trace property in phi}). At the price of an increase in the number
of components from $d^{3}$ (for $d^2$ vectors in $\mathbb{C}^{d}$) 
                to $d^{4}$ (for $d^2$ vectors in $\mathbb{C}^{d^{2}}$), we 
have got rid of the square
modulus to result in a single scalar product (compare Eqs.~(\ref{Result 1})
and (\ref{Trace property in phi})), what may prove to be suitable for
another way to search for SIC-POVMs. Moreover, our relation (\ref{Result 1})
is independent of any hypothesis on the rank of the operators $P_{x}$. In
fact, there exists a lot of relations among these $d^{4}$ coefficients that
decrease the effective number of coefficients to be found and give structural
constraints on them. Those relations are highly sensitive to the choice of
the basis $\{u_{i}:i=1,2,\cdots,d^{2}\}$\ and we are going to exhibit an
example of such a set of relations by choosing the basis to consist of Racah
unit tensors.

\bigskip

The cornerstone of this approach is to identify $\mathbb{C}^{d}$ with a
subspace $\varepsilon (j)$ of constant angular momentum $j=(d-1)/2$. Such a
subspace is spanned by the set $\{|j,m\rangle :m=-j,-j+1,\cdots,j\}$, where 
$|j,m\rangle $ is an eigenvector of the square and the $z$-component of a
generalized angular momentum operator. Let $\mathbf{u}^{(k)}$ be the Racah
unit tensor \cite{23Racah} of order $k$ (with $k=0,1,\cdots,2j$) defined by
its $2k+1$ components ${u}_{q}^{(k)}$ (where $q=-k,-k+1,\cdots,k$) through 
\begin{equation}
u_{q}^{(k)}=\sum_{m=-j}^{j}\sum_{m^{\prime }=-j}^{j}(-1)^{j-m}\left( 
\begin{array}{ccc}
j & k & j \\ 
-m & q & m^{\prime }%
\end{array}%
\right) |j,m\rangle \langle j,m^{\prime }|,  \label{def de ukq}
\end{equation}
where $\left( \cdots \right) $ denotes a 3--$jm$ Wigner symbol. For fixed $j$%
, the $(2j+1)^{2}$ operators ${u}_{q}^{(k)}$ (with $k = 0, 1, \cdots, 2j$ and 
$q=-k, -k+1, \cdots, k$) act on $\varepsilon (j)\sim \mathbb{C}^{d}$ and form 
a basis of the Hilbert space $\mathbb{C}^{N}$ of dimension $N=(2j+1)^{2}$, the
inner product in $\mathbb{C}^{N}$ being the Hilbert--Schmidt product. The
formulas (involving unit tensors, 3--$jm$ and 6--$j$ symbols) relevant for
this work are given in Appendix (see also \cite{23Racah} to 
\cite{25Kibler M}). We must remember that those Racah operators are not
normalized to unity (see relation (\ref{HS pour u})). So this will generate
an extra factor when defining $v_{i}(x)$.

Each operator $P_{x}$ can be developed as a linear combination of the
operators ${u}_{q}^{(k)}$. Hence, we have
\begin{equation}
P_{x}=\sum_{k=0}^{2j}\sum_{q=-k}^{k}c_{kq}(x)u_{q}^{(k)},  \label{P en u}
\end{equation}
where the unknown expansion coefficients $c_{kq}(x)$ are \textit{a priori}
complex numbers. The determination of the operators $P_{x}$ is thus
equivalent to the determination of the coefficients $c_{kq}(x)$, which are
formally given by
\begin{equation}
c_{kq}(x)=(2k+1)\overline{\langle \Phi _{x}|{u}_{q}^{(k)}|\Phi _{x}\rangle },
\end{equation}
as can be seen by multiplying each member of Eq.~(\ref{P en u}) by the 
adjoint of ${u}_{p}^{(\ell )}$ and then using Eq.~(\ref{HS pour u}) of Appendix.

By defining the vector 
\begin{equation}
v(x)=(v_{1}(x),v_{2}(x),\cdots,v_{N}(x)),\quad N=(2j+1)^{2}
\end{equation}
via 
\begin{equation}
v_{i}(x)=\frac{1}{\sqrt{2k+1}}c_{kq}(x),\quad i=k^{2}+k+q+1,
\end{equation}
the following properties and relations are obtained.

\begin{itemize}
\item \textit{The first component }$v_{1}(x)$\textit{\ of }$v(x)$\textit{\
does not depend on }$x$ since
\begin{equation}
c_{00}(x)=\frac{1}{\sqrt{2j+1}}
\end{equation}
for all $x\in \left\{ 1,2,\cdots,(2j+1)^{2}\right\} $.

Proof: Take the trace of Eq.~(\ref{P en u}) and use Eq.~(\ref{BCR}) of
Appendix.

\item The components $v_{i}(x)$ of $v(x)$ satisfy the \textit{complex
conjugation property} described by 
\begin{equation}
\overline{c_{kq}(x)}=(-1)^{q}c_{k-q}(x)
\end{equation}
for all $x\in \left\{ 1,2,\cdots,(2j+1)^{2}\right\} $, $k\in \left\{
0,1,\cdots,2j\right\} $ and $q\in \left\{ -k,-k+1,\cdots,k\right\} $.

Proof: Use the Hermitian property of $P_{x}$ and Eq.~(\ref{HC de u}) of
Appendix.

\item In terms of $c_{kq}$, Eq.~(\ref{Result 1}) reads
\begin{equation}
\sum_{k=0}^{2j}\frac{1}{2k+1}\sum_{q=-k}^{k}\overline{c_{kq}(x)}c_{kq}(y)=
\frac{1}{2(j+1)}\left[ (2j+1)\delta _{x,y}+1\right]
\end{equation}

for all $x, y \in \left\{ 1, 2,    \cdots, (2j+1)^{2}\right\}$, 
where the sum over $q$ is $\mathrm{SO}(3)$ rotationally invariant.

Proof: The proof is trivial.

\item The coefficients $c_{kq}(x)$ are solutions of the 
\textit{nonlinear system} given by 
\begin{eqnarray}
\frac{1}{2K+1}c_{KQ}(x) &=&(-1)^{2j-Q}\sum_{k=0}^{2j}\sum_{\ell
=0}^{2j}\sum_{q=-k}^{k}\sum_{p=-\ell }^{\ell }\left( 
\begin{array}{ccc}
k & \ell  & K \\ 
-q & -p & Q%
\end{array}%
\right)   \nonumber \\
&\times &\left\{ 
\begin{array}{ccc}
k & \ell  & K \\ 
j & j & j%
\end{array}%
\right\} c_{kq}(x)c_{\ell p}(x)  \label{NLS ckq}
\end{eqnarray}%
for all $x$ $\in$ $\{ 1, 2,    \cdots, (2j+1)^{2}\}$, 
        $K$ $\in$ $\{ 0, 1,    \cdots, 2j        \}$ and 
	$Q$ $\in$ $\{-K, -K+1, \cdots, K         \}$.

Proof: Consider $P_{x}^{2}=P_{x}$ and use the coupling relation (\ref{uu})
of Appendix involving a 3--$jm$ and a 6--$j$ Wigner symbols.

As a corollary of the latter property, by taking $K=0$ and using Eqs.~(\ref{3j0})
and (\ref{6j0}) of Appendix, we get again the normalization relation 
$\Vert v(x) \Vert ^{2}=v(x)\cdot v(x)=1$.

\item All coefficients $c_{kq}(x)$ are connected through the \textit{sum rule} 
\begin{equation}
\sum_{x=1}^{(2j+1)^{2}}\sum_{k=0}^{2j}\sum_{q=-k}^{k}c_{kq}(x)\left( 
\begin{array}{ccc}
j & k & j \\ 
-m & q & m^{\prime }%
\end{array}%
\right) =(-1)^{j-m}(2j+1)\delta _{m,m^{\prime }},
\end{equation}
which turns out to be useful for global checking purposes.

Proof: Take the $jm$--$jm^{\prime }$ matrix element of the resolution of the
identity in terms of the operators $P_{x}/(2j+1)$.
\end{itemize}

\section{MUBs}

A complete set of MUBs in the Hilbert space $\mathbb{C}^{d}$ is a set of 
$d(d+1)$ vectors $|a\alpha \rangle \in \mathbb{C}^{d}$ such that 
\begin{equation}
|\langle a\alpha |b\beta \rangle |^{2}=\delta _{\alpha ,\beta }\delta _{a,b}+%
\frac{1}{d}(1-\delta _{a,b}),  \label{def MUBs}
\end{equation}
where $a=0,1,\cdots,d$ and $\alpha =0,1,\cdots,d-1$. The indices of type 
$a$ refer to the bases and, for fixed $a$, the index $\alpha $ refers to one
of the $d$ vectors of the basis corresponding to $a$. We know that such a
complete set exists if $d$ is a prime or the power of a prime (\textit{e.g.}, 
see \cite{13Delsarte}-\cite{19Klappen}).

The approach developed in Sec.~2 for SIC-POVMs can be applied to MUBs too.
Let us suppose that it is possible to find $d+1$ sets $S_{a}$ (with 
$a=0,1,\cdots,d$) of vectors in $\mathbb{C}^{d}$, each set $S_{a}=\{|a\alpha
\rangle :\alpha =0,1,\cdots,d-1\}$ containing $d$ vectors $|a\alpha \rangle 
$ such that Eq.~(\ref{def MUBs}) be satisfied. This amounts to finding 
$d(d+1)$ \textit{projection operators} 
\begin{equation}
\Pi _{a\alpha }=|a\alpha \rangle \langle a\alpha |
\end{equation}
satisfying the \textit{trace condition}
\begin{equation}
{\rm Tr}\left( \Pi _{a\alpha }\Pi _{b\beta }\right) =\delta
_{\alpha ,\beta }\delta _{a,b}+\frac{1}{d}(1-\delta _{a,b}),
\end{equation}
where the trace is taken on $\mathbb{C}^{d}$. Therefore, they also form a 
\textit{nonorthogonal decomposition of the identity}
\begin{equation}
\frac{1}{d+1}\sum_{a=0}^{d}\sum_{\alpha =0}^{d-1}\Pi _{a\alpha }=\mathbb{I}.
\end{equation}
As in Sec.~2, we develop each operator $\Pi _{a\alpha }$\ on an orthonormal
basis with expansion coefficients $w_{i}(a\alpha )$. Thus we get vectors $%
w(a\alpha )$\ in $\mathbb{C}^{d^{2}}$
\begin{equation}
w(a\alpha )=\left( w_{1}(a\alpha ),w_{2}(a\alpha ),\cdots,w_{d^{2}}(a\alpha
)\right) 
\end{equation}
such that%
\begin{equation}
w(a\alpha )\cdot w(b\beta )=\delta _{\alpha ,\beta }\delta _{a,b}+\frac{1}{d}%
(1-\delta _{a,b})  \label{Result 2}
\end{equation}
for all $a,b\in \left\{ 0,1,\cdots,d\right\} $ and $\alpha ,\beta \in
\left\{ 0,1,\cdots,d-1\right\} $.

\bigskip

Now we draw the same relations as for POVMs by choosing the Racah operators
to be our basis in $\mathbb{C}^{d^{2}}$. We assume once again that the
Hilbert space $\mathbb{C}^{d}$ is realized by $\varepsilon (j)$ with $%
j=(d-1)/2$. Then, each operator $\Pi _{a\alpha }$ can be developed on the
basis of the $(2j+1)^{2}$ operators ${u}_{q}^{(k)}$ as 
\begin{equation}
\Pi _{a\alpha }=\sum_{k=0}^{2j}\sum_{q=-k}^{k}d_{kq}(a\alpha )u_{q}^{(k)},
\end{equation}
to be compared with Eq.~(\ref{P en u}). The expansion coefficients are%
\begin{equation}
d_{kq}(a\alpha )=(2k+1)\overline{\langle a\alpha |{u}_{q}^{(k)}|a\alpha
\rangle }  \label{dkq}
\end{equation}
for all 
$a$      $\in$ $\{0,    1,\cdots,2j+1\}$, 
$\alpha$ $\in$ $\{0,    1,\cdots,2j\}$, 
$k$      $\in$ $\{0,    1,\cdots,2j\}$ and 
$q$      $\in$ $\{-k,-k+1,\cdots,k\}$. For $a$ and $\alpha$ fixed, the
complex coefficients $d_{kq}(a\alpha )$ define a vector 
\begin{equation}
w(a\alpha )=\left( w_{1}(a\alpha ),w_{2}(a\alpha ),\cdots,w_{N}(a\alpha
)\right) ,\quad N=(2j+1)^{2}
\end{equation}
in the Hilbert space $\mathbb{C}^{N}$, the components of which are given by 
\begin{equation}
w_{i}(a\alpha )=\frac{1}{\sqrt{2k+1}}d_{kq}(a\alpha ),\quad i=k^{2}+k+q+1.
\end{equation}
We are thus led to the following properties and relations. The proofs are
similar to those in Sec.~2.

\begin{itemize}
\item \textit{First component }$w_{1}(a\alpha )$\textit{\ of }$w(a\alpha)$:
\begin{equation}
d_{00}(a\alpha )=\frac{1}{\sqrt{2j+1}}
\end{equation}
for all $a\in \left\{ 0,1,\cdots,2j+1\right\} $ and $\alpha \in \left\{
0,1,\cdots,2j\right\} $.

\item \textit{Complex conjugation property}: 
\begin{equation}
\overline{d_{kq}(a\alpha )}=(-1)^{q}d_{k-q}(a\alpha )
\end{equation}
for all $a\in \left\{ 0,1,\cdots,2j+1\right\} $, $\alpha \in \left\{
0,1,\cdots,2j\right\} $, $k\in \left\{ 0,1,\cdots,2j\right\} $ and $q\in
\left\{ -k,-k+1,\cdots,k\right\} $.

\item \textit{Rotational invariance}:%
\begin{equation}
\sum_{k=0}^{2j}\frac{1}{2k+1}\sum_{q=-k}^{k}\overline{d_{kq}(a\alpha )}
d_{kq}(b\beta )=\delta _{\alpha ,\beta }\delta _{a,b}+\frac{1}{2j+1}
(1-\delta _{a,b})
\end{equation}
for all $a,b           \in \left\{ 0,1,\cdots,2j+1\right\}$ and 
        $\alpha ,\beta \in \left\{ 0,1,\cdots,2j\right\}$.

\item \textit{Tensor product formula}: 
\begin{eqnarray}
\frac{1}{2K+1}d_{KQ}(a\alpha ) &=&(-1)^{2j-Q}\sum_{k=0}^{2j}\sum_{\ell
=0}^{2j}\sum_{q=-k}^{k}\sum_{p=-\ell }^{\ell }\left( 
\begin{array}{ccc}
k & \ell  & K \\ 
-q & -p & Q%
\end{array}%
\right)   \nonumber \\
&\times &\left\{ 
\begin{array}{ccc}
k & \ell  & K \\ 
j & j & j%
\end{array}%
\right\} d_{kq}(a\alpha )d_{\ell p}(a\alpha )  \label{NLS dkq}
\end{eqnarray}%
for all $a\in \left\{ 0,1,\cdots,2j+1\right\} $, $\alpha \in \left\{
0,1,\cdots,2j\right\} $, $K\in \left\{ 0,1,\cdots,2j\right\} $ and $Q\in
\left\{ -K,-K+1,\cdots,K\right\} $.

\item \textit{Sum rule}: 
\begin{equation}
\sum_{a=0}^{2j+1}\sum_{\alpha
=0}^{2j}\sum_{k=0}^{2j}\sum_{q=-k}^{k}d_{kq}(a\alpha )\left( 
\begin{array}{ccc}
j & k & j \\ 
-m & q & m^{\prime }%
\end{array}%
\right) =(-1)^{j-m}2(2j+1)\delta _{m,m^{\prime }}
\end{equation}
which involves all coefficients $d_{kq}(a\alpha )$.
\end{itemize}

\section{CONCLUSIONS}

Although the structure of the relations in Sec.~1 on the one hand and Sec.~2
on the other hand is very similar, there are deep differences between the
two sets of results. The similarities are reminiscent of the fact that both
MUBs and SIC-POVMs can be linked to finite affine planes 
\cite{10Grassl,102Grassl,17SPR,18Bengtsson,oott} 
and to complex projective 2--designs 
\cite{06Zauner,08Renes,16Barnum,19Klappen}. On the other side, there are two
arguments in favor of the differences between relations (\ref{Result 1}) and
(\ref{Result 2}). First, the problem of constructing SIC-POVMs in dimension 
$d$ is not equivalent to the existence of an affine plane of order $d$ 
\cite{10Grassl,102Grassl}. Second, there is a consensus around the conjecture 
according to which there exists a complete set of MUBs in dimension $d$ if 
and only if there exists an affine plane of order $d$ \cite{17SPR}.

In dimension $d$, to find $d^{2}$ operators $P_{x}$ of a SIC-POVM acting on
the Hilbert space $\mathbb{C}^{d}$ amounts to find $d^{2}$ vectors $v(x)$ in
the Hilbert space $\mathbb{C}^{N}$ with $N=d^{2}$ satisfying 
\begin{equation}
\Vert v_{x}\Vert =1,\quad v(x)\cdot v(y)=\frac{1}{d+1}\ \mathrm{for}\ x\not=y
\label{v.v de conclu}
\end{equation}
(the norm $\Vert v(x) \Vert$ of each vector $v(x)$ is 1 and the 
angle $\omega _{xy}$ of any pair of vectors $v(x)$ and $v(y)$ is 
$\omega _{xy}=\cos^{-1}[1/(d+1)]$ for $x\not=y$).

In a similar way, to find $d+1$ MUBs of $\mathbb{C}^{d}$ is equivalent to
find $d+1$ sets $S_{a}$ (with $a=0,1,\cdots,d$) of $d$ vectors, 
\textit{i.e.}, $d(d+1)$ vectors in all, $w(a\alpha )$ in 
$\mathbb{C}^{N}$ with $N=d^{2}$ satisfying 
\begin{equation}
w(a\alpha )\cdot w(a\beta )=\delta _{\alpha ,\beta },\quad w(a\alpha )\cdot
w(b\beta )=\frac{1}{d}\ \mathrm{for}\ a\not=b  \label{w.w de conclu}
\end{equation}
(each set $S_{a}$ consists of $d$ orthonormalized vectors and the angle 
$\omega _{a\alpha b\beta }$ of any vector $w(a\alpha )$ of a set $S_{a}$ with
any vector $w(b\beta )$ of a set $S_{b}$ is $\omega _{a\alpha b\beta }=\cos
^{-1}(1/d)$ for $a\not=b$).

According to a well accepted conjecture \cite{06Zauner,08Renes}, SIC-POVMs
should exist in any dimension. The present study shows that in order to
prove this conjecture it is sufficient to prove that Eq.~(\ref{v.v de conclu}) 
admits solutions for any value of $d$.

The situation is different for MUBs. In dimension $d$, it is known that
there exist $d+1$ sets of $d$ vectors of type $|a\alpha \rangle $ in $%
\mathbb{C}^{d}$ satisfying Eq.~(\ref{def MUBs}) when $d$ is a prime or the
power of a prime. This shows that Eq.~(\ref{w.w de conclu}) can be solved
for $d$ prime or power of a prime. For $d$ prime, it is possible to find an
explicit solution of Eq.~(\ref{def MUBs}). In fact, we have 
\cite{20Kibler Planat,21Albouy} 
\begin{eqnarray}
|a\alpha \rangle  &=&\frac{1}{\sqrt{2j+1}}\sum_{m=-j}^{j}\omega
^{(j+m)(j-m+1)a/2+(j+m)\alpha }|j,m\rangle , \\
\omega  &=&\exp \left( \mathrm{i}\frac{2\pi }{2j+1}\right) ,\quad j=\frac{1}{%
2}(d-1)
\end{eqnarray}%
for $a,\alpha \in \left\{ 0,1,\cdots,2j\right\} $ while%
\begin{equation}
|a\alpha \rangle =|j,m\rangle 
\end{equation}
for $a=2j+1$ and $\alpha =j+m=0,1,\cdots,2j$. Then, Eq.~(\ref{dkq}) yields 
\begin{eqnarray}
d_{kq}(a\alpha ) &=&\frac{2k+1}{2j+1}\sum_{m=-j}^{j}\sum_{m^{\prime
}=-j}^{j}\omega ^{\theta (m,m^{\prime })}(-1)^{j-m}\left( 
\begin{array}{ccc}
j & k & j \\ 
-m & q & m^{\prime }%
\end{array}%
\right) , \\
\theta (m,m^{\prime }) &=&(m-m^{\prime })\left[ \frac{1}{2}(1-m-m^{\prime
})a+\alpha \right]   \label{solution}
\end{eqnarray}%
for $a,\alpha \in \left\{ 0,1,\cdots,2j\right\} $ while 
\begin{equation}
d_{kq}(a\alpha )=\delta _{q,0}(2k+1)(-1)^{j-m}\left( 
\begin{array}{ccc}
j & k & j \\ 
-m & 0 & m%
\end{array}%
\right)   \label{solutionprime}
\end{equation}
for $a=2j+1$ and $\alpha = j+m = 0, 1, \cdots, 2j$. It can be shown that 
Eqs.~(\ref{solution}) and (\ref{solutionprime}) are in agreement with the results
of Sec.~3. We thus have a solution of the equations for the results of
Sec.~3 when $d$ is prime. As an open problem, it would be worthwhile to find
an explicit solution for the coefficients $d_{kq}(a\alpha )$ when $d=2j+1$
is any positive power of a prime. Finally, note that to prove (or disprove)
the conjecture according to which a complete set of MUBs in dimension $d$
exists only if $d$ is a prime or the power of a prime is equivalent to prove
(or disprove) that Eq.~(\ref{w.w de conclu}) has a solution only if $d$ is a
prime or the power of a prime.

\section*{APPENDIX: WIGNER-RACAH ALGEBRA OF $\mathrm{SU}(2) \supset \mathrm{U%
}(1)$}

We limit ourselves to those basic formulas for the Wigner-Racah algebra of
the chain $\mathrm{SU}(2)\supset \mathrm{U}(1)$ which are necessary to
derive the results of this paper. The summations in this appendix have to be
extended to the allowed values for the involved magnetic and angular
momentum quantum numbers.

The definition (\ref{def de ukq}) of the components $u_{q}^{(k)}$ of the
Racah unit tensor $\mathbf{u}^{(k)}$ yields 
\begin{equation}
\langle j,m|u_{q}^{(k)}|j,m^{\prime }\rangle =(-1)^{j-m}\left( 
\begin{array}{ccc}
j & k & j \\ 
-m & q & m^{\prime }%
\end{array}%
\right) ,  \label{def standard de ukq}
\end{equation}
from which we easily obtain the Hermitian conjugation property 
\begin{equation}
{u_{q}^{(k)}}^{\dagger }=(-1)^{q}u_{-q}^{(k)}.  \label{HC de u}
\end{equation}
The 3--$jm$ Wigner symbol in Eq.~(\ref{def standard de ukq}) satisfies the
orthogonality relations 
\begin{equation}
\sum_{mm^{\prime }}\left( 
\begin{array}{ccc}
j & j^{\prime } & k \\ 
m & m^{\prime } & q%
\end{array}%
\right) \left( 
\begin{array}{ccc}
j & j^{\prime } & \ell  \\ 
m & m^{\prime } & p%
\end{array}%
\right) =\frac{1}{2k+1}\delta _{k,\ell }\delta _{q,p}\Delta (j,j^{\prime },k)
\label{orthogo mm'}
\end{equation}
and 
\begin{equation}
\sum_{kq}(2k+1)\left( 
\begin{array}{ccc}
j & j^{\prime } & k \\ 
m & m^{\prime } & q%
\end{array}%
\right) \left( 
\begin{array}{ccc}
j & j^{\prime } & k \\ 
M & M^{\prime } & q%
\end{array}%
\right) =\delta _{m,M}\delta _{m^{\prime },M^{\prime }}.
\end{equation}
The trace relation on the space $\varepsilon (j)$
\begin{equation}
{\rm Tr}\left( {u_{q}^{(k)}}^{\dagger }u_{p}^{(\ell )}\right) =
\frac{1}{2k+1}\delta _{k,\ell }\delta _{q,p}\Delta (j,j,k)
\label{HS pour u}
\end{equation}
easily follows by combining Eqs.~(\ref{def standard de ukq}) and (\ref%
{orthogo mm'}). Furthermore, by introducing 
\begin{equation}
\left( 
\begin{array}{ccc}
j & j^{\prime } & 0 \\ 
m & -m^{\prime } & 0%
\end{array}%
\right) =\delta _{j,j^{\prime }}\delta _{m,m^{\prime }}(-1)^{j-m}\frac{1}{%
\sqrt{2j+1}}  \label{3j0}
\end{equation}
in Eq.~(\ref{orthogo mm'}), we obtain the sum rule
\begin{equation}
\sum_{m}(-1)^{j-m}\left( 
\begin{array}{ccc}
j & k & j \\ 
-m & q & m%
\end{array}%
\right) =\sqrt{2j+1}\delta _{k,0}\delta _{q,0}\Delta (j,k,j),  \label{BCR}
\end{equation}
known in spectroscopy as the barycenter theorem.

There are several relations involving 3--$jm$ and 6--$j$ symbols.
In particular, we have 
\begin{eqnarray}
\sum_{mm^{\prime }M}  &(-1)^{j-M}& \left( 
\begin{array}{ccc}
j & k & j \\ 
-m & q & M%
\end{array}%
\right) \left( 
\begin{array}{ccc}
j & \ell  & j \\ 
-M & p & m^{\prime }%
\end{array}%
\right) \left( 
\begin{array}{ccc}
j & K & j \\ 
-m & Q & m^{\prime }%
\end{array}%
\right)   \nonumber \\
                   = &(-1)^{2j-Q}& \left( 
\begin{array}{ccc}
k & \ell  & K \\ 
-q & -p & Q%
\end{array}%
\right) \left\{ 
\begin{array}{ccc}
k & \ell  & K \\ 
j & j & j%
\end{array}%
\right\} ,  \label{3CG=1CGxW}
\end{eqnarray}%
where $\left\{ \cdots \right\} $\ denotes a 6--$j$ Wigner symbol (or $%
\overline{W}$\ Racah coefficient). Note that the introduction of 
\begin{equation}
\left\{ 
\begin{array}{ccc}
k & \ell  & 0 \\ 
j & j & J%
\end{array}%
\right\} =\delta _{k,\ell }(-1)^{j+k+J}\frac{1}{\sqrt{(2k+1)(2j+1)}}
\label{6j0}
\end{equation}
in Eq.~(\ref{3CG=1CGxW}) gives back Eq.~(\ref{orthogo mm'}). Equation (\ref%
{3CG=1CGxW}) is central in the derivation of the coupling relation 
\begin{equation}
u_{q}^{(k)}u_{p}^{(\ell )}=\sum_{KQ}(-1)^{2j-Q}(2K+1)\left( 
\begin{array}{ccc}
k & \ell  & K \\ 
-q & -p & Q%
\end{array}%
\right) \left\{ 
\begin{array}{ccc}
k & \ell  & K \\ 
j & j & j%
\end{array}%
\right\} u_{Q}^{(K)}.  \label{uu}
\end{equation}
Equation (\ref{uu}) makes it possible to calculate the commutator $%
[u_{q}^{(k)},u_{p}^{(\ell )}]$ which shows that the set 
$\{u_{q}^{(k)} : k = 0, 1, \cdots, 2j; \quad q=-k,-k+1,\cdots,k\}$ 
can be used to
span the Lie algebra of the unitary group U($2j+1$). The latter result is at
the root of the expansions (\ref{NLS ckq}) and (\ref{NLS dkq}).

\bigskip

\noindent \textbf{Note added in version 3}

After the submission of the present paper for publication in 
\textit{Journal of Russian Laser Research},
a pre-print dealing with the existence of SIC-POVMs was posted
on arXiv \cite{32Hall}. The main result in \cite{32Hall} is that 
SIC-POVMs exist in all dimensions. As a corollary of this 
result, Eq. (34) admits solutions in any dimension.

\bigskip

\noindent \textbf{Acknowledgements}

\bigskip

This work was presented at the \textit{International Conference on
Squeezed States and Uncertainty Relations}, University of Bradford, England 
(ICSSUR'07). The authors wish to thank the organizer A. Vourdas 
and are grateful to D. M. Appleby, V. I. Man'ko and M. Planat for interesting 
comments.



\newpage

\begin{thebibliography}{99}
\bibitem{01Peres} A. Peres, ``Quantum Theory: Concepts and Methods'', 
\textit{Dordrecht: Kluwer} (1995)

\bibitem{02Vourdas} A. Vourdas, \textit{J. Phys. A: Math. Gen.} \textbf{38},
8453 (2005)

\bibitem{add Muynck} W. M. de Muynck, ``Foundations of Quantum Mechanics, an
Empiricist Approach'', 
\textit{Dordrecht: Kluwer} (2002)

\bibitem{03Jauch} J. M. Jauch and C. Piron, \textit{Helv. Phys. Acta} 
\textbf{40}, 559 (1967)

\bibitem{04Davies1} E. B. Davies and J. T. Levis, \textit{Comm. Math. Phys.} 
\textbf{17}, 239 (1970)

\bibitem{05Davies2} E. B. Davies, \textit{IEEE Trans. Inform. Theory} 
\textbf{IT-24}, 596 (1978)

\bibitem{add Kraus} K. Kraus, ``States, Effects, and Operations'', 
\textit{Lect. Notes Phys.} \textbf{190} (1983)

\bibitem{06Zauner} G. Zauner, \textit{Diploma Thesis}, University of Wien
(1999)

\bibitem{07Caves} C. M. Caves, C. A. Fuchs and R. Schack, \textit{J. Math.
Phys.} \textbf{43}, 4537 (2002)

\bibitem{08Renes} J. M. Renes, R. Blume-Kohout, A. J. Scott and C. M. Caves, 
\textit{J. Math. Phys.} \textbf{45}, 2171 (2004)

\bibitem{09Appleby} D. M. Appleby, \textit{J. Math. Phys.} \textbf{46},
052107 (2005)

\bibitem{10Grassl} M. Grassl, \textit{Proc. ERATO Conf. Quant. Inf. Science
(EQIS 2004)} ed. J. Gruska, Tokyo (2005)

\bibitem{102Grassl} M. Grassl, \textit{Elec. Notes Discrete Math.} 
\textbf{20}, 151 (2005)

\bibitem{11Weigert} S. Weigert, \textit{Int. J. Mod. Phys.} B \textbf{20}, 1942 (2006)

\bibitem{12Schwinger} J. Schwinger, \textit{Proc. Nat. Acad. Sci. USA} 
\textbf{46}, 570 (1960)

\bibitem{13Delsarte} P. Delsarte, J. M. Goethals and J. J. Seidel, 
\textit{Philips Res. Repts.} \textbf{30}, 91 (1975)

\bibitem{14Ivanovic} I. D. Ivanovi\'{c}, \textit{J. Phys. A: Math. Gen.} 
\textbf{14}, 3241 (1981)

\bibitem{15Wootters} W. K. Wootters, \textit{Ann. Phys. (N.Y.)} 
\textbf{176}, 1 (1987)

\bibitem{16Barnum} H. Barnum, \textit{Preprint} quant-ph/0205155 (2002)

\bibitem{addBandyo} S. Bandyopadhyay, P. O. Boykin, V. Roychowdhury and F. Vatan, 
\textit{Algorithmica} \textbf{34}, 512 (2002)

\bibitem{add Pitt. Rub. 1} A. O. Pittenger and M. H. Rubin, 
\textit{Linear Alg. Appl.} \textbf{390}, 255 (2004)

\bibitem{17SPR} M. Saniga, M. Planat and H. Rosu, \textit{J. Opt. B: Quantum
Semiclassical Opt.} \textbf{6}, L19 (2004)

\bibitem{18Bengtsson} I. Bengtsson and \AA . Ericsson, \textit{Open Syst.
Inf. Dyn.} \textbf{12}, 107 (2005)

\bibitem{19Klappen} A. Klappenecker and M. R\"{o}tteler, \textit{Preprint}
quant-ph/0502031 (2005)

\bibitem{oott} W. K. Wootters, \textit{Found. Phys.} \textbf{36}, 112 (2006)

\bibitem{20Kibler Planat} M. R. Kibler and M. Planat, \textit{Int. J. Mod.
Phys.} B \textbf{20}, 1802 (2006)

\bibitem{21Albouy} O. Albouy and M. R. Kibler, \textit{SIGMA} \textbf{3},
article 076 (2007)

\bibitem{22Breuer} H.-P. Breuer, \textit{J. Phys. A: Math. Gen.} 
\textbf{38}, 9019 (2005)

\bibitem{23Racah} G. Racah, \textit{Phys. Rev.} \textbf{62}, 438 (1942)

\bibitem{24Fano} U. Fano and G. Racah, ``Irreducible Tensorial Sets'', 
\textit{New York: Academic} (1959)

\bibitem{25Kibler M} M. Kibler and G. Grenet, \textit{J. Math. Phys.} 
\textbf{21}, 422 (1980)

\bibitem{32Hall} J.L. Hall and A. Rao, \textit{Preprint} quant-ph/0707.3002v1 
(20 July 2007)

\end{thebibliography}
\end{document}